# Identification of a Turnover in the Initial Mass Function of a Young Stellar Cluster Down to 0.5 $M_J$

Matthew De Furio[1,2], Michael R. Meyer[3], Thomas Greene[4], Klaus Hodapp[5], Doug Johnstone[6,7], Jarron Leisenring[8], Marcia Rieke[8], Massimo Robberto[9,10], Thomas Roellig[4], Gabriele Cugno[11], Eleonora Fiorellino[12,13], Carlo F. Manara[14], Roberta Raileanu[15], and Sierk van Terwisga[16]

[1] Department of Astronomy, The University of Texas at Austin, 2515 Speedway, Stop C1400, Austin, TX 78712, USA; defurio@utexas.edu
[2] NSF Astronomy and Astrophysics Postdoctoral Fellow
[3] Department of Astronomy, University of Michigan, Ann Arbor, MI 48109, USA
[4] MS 245-6, NASA Ames Research Center, Moffett Field, CA 94035, USA
[5] University of Hawaii, Hilo, HI 96720, USA
[6] NRC Herzberg Astronomy and Astrophysics, 5071 West Saanich Rd, Victoria, BC, V9E 2E7, Canada
[7] Department of Physics and Astronomy, University of Victoria, Victoria, BC, V8P 5C2, Canada
[8] Steward Observatory, University of Arizona, Tucson, AZ 85721, USA
[9] Space Telescope Science Institute, 3700 San Martin Dr., Baltimore, MD 21218, USA
[10] Johns Hopkins University, 3400 N. Charles St., Baltimore, MD 21218, USA
[11] Department of Astrophysics, University of Zurich, Winterthurerstr. 90 CH-8057 Zurich, Switzerland
[12] Instituto de Astrofísica de Canarias, IAC, Vía Láctea s/n, 38205 La Laguna (S.C.Tenerife), Spain
[13] Departamento de Astrofísica, Universidad de La Laguna, 38206 La Laguna (S.C.Tenerife), Spain
[14] European Southern Observatory, Karl-Schwarzschild-Strasse 2, 85748 Garching bei Munchen, Germany
[15] Department of Computer Science, University College London, London, UK
[16] Space Research Institute, Austrian Academy of Sciences, Schmiedlstr. 6, 8042 Graz, Austria

## Abstract

A successful theory of star formation should predict the number of objects as a function of their mass produced through star-forming events. Previous studies in star-forming regions and the solar neighborhood have identified a mass function increasing from the hydrogen-burning limit down to about 10 $M_J$. Theory predicts a limit to the fragmentation process, providing a natural turnover in the mass function down to the opacity limit of turbulent fragmentation, thought to be near 1–10 $M_J$. Programs to date have not been sensitive enough to probe the hypothesized opacity limit of fragmentation. We present the first identification of a turnover in the initial mass function below 12 $M_J$ within NGC 2024, a young star-forming region. With JWST/NIRCam deep exposures across 0.7–5 $\mu$m, we identified several free-floating objects down to roughly 3 $M_J$ with sensitivity to 0.5 $M_J$. We present evidence for a double power-law model increasing from about 60 $M_J$ to roughly 12 $M_J$, consistent with previous studies, followed by a decrease down to 0.5 $M_J$. Our results support the predictions of star and brown dwarf formation theory, identifying the theoretical turnover in the mass function and suggesting the fundamental limit of turbulent fragmentation to be near 3 $M_J$.

## 1. Introduction

The initial mass function (IMF), the distribution of the number of objects in a population as a function of mass, is a fundamental outcome of the star formation process and one of the most commonly studied to explore the star formation process. The canonical IMF is a power law above 1 $M_\odot$ (E. E. Salpeter 1955), with variations including a log-normal distribution (G. Chabrier 2003) or a broken power law (P. Kroupa 2001) down to substellar masses.

Many surveys have attempted to characterize the mass function down to Jupiter-mass scales to explore the fundamental limit of turbulent fragmentation in the Galactic field (e.g., W. M. J. Best et al. 2024; J. D. Kirkpatrick et al. 2024), the halo, star clusters (e.g., E. Moraux et al. 2004), and young star-forming regions (e.g., M. Gennaro & M. Robberto 2020). Early studies in the Galactic field modeled the mass function as a broken power law ($dN \propto m^{-\alpha}$) and calculated a power-law index over 0.01–0.08 $M_\odot$ as $\alpha = 0.3 \pm 0.7$ (P. Kroupa 2001). More recent work characterized the mass function using stars and brown dwarfs within 20 pc and found $\alpha = 0.6 \pm 0.1$ for 0.01–0.05 $M_\odot$ (J. D. Kirkpatrick et al. 2021, 2024). Microlensing surveys are sensitive to free-floating super-Earths, and they found a mass function with a power-law index of $\alpha = 0.8$ with a $3\sigma$ confidence interval of [0.2, 1.3] for masses 0.01–0.08 $M_\odot$ (P. Mróz et al. 2017). However, none of these programs were able to characterize the IMF, the product of the star and brown dwarf formation process, below 0.01 $M_\odot$.

With the launch of the James Webb Space Telescope (JWST), multiple programs were designed to detect free-floating planetary-mass objects in young star-forming regions, with the ultimate goal of identifying the opacity limit for fragmentation, i.e., the end of the IMF, hypothesized to be 1–10 $M_J$. While not characterizing the mass function, a NIRCam/NIRSpec program in IC 348 identified evidence for a 3–4 $M_J$ cluster member with no lower-mass objects identified

(K. L. Luhman et al. 2024). Similarly, a NIRISS study in NGC 1333 identified free-floating planetary-mass objects with estimated masses of 5–15 $M_J$ without any observed lower-mass sources, despite sensitivity to such objects (A. B. Langeveld et al. 2024). A recent study in the ONC claims to detect planetary-mass objects with masses down to 0.6 $M_J$ (M. J. McCaughrean & S. G. Pearson 2023), in contrast to the results in IC 348 and NGC 1333. Yet, contamination from Galactic and extragalactic sources may bias their results as indicated by K. L. Luhman (2024).

In this Letter, we present the first results of the JWST/NIRCam (M. J. Rieke et al. 2023) program GTO-1190 (PI: M. Meyer), exploring the IMF down to sub-Jupiter masses in NGC 2024, a young embedded star-forming region. In Section 2, we describe the observations, In Section 3, we describe the automated process used to detect likely planetary-mass cluster members. In Section 4, we present our analysis of the mass function and the results of our survey in NGC 2024. In Section 5, we discuss the implications of our work. In Section 6, we provide concluding remarks.

## 2. Observations

The data used in this Letter were observed by the Near Infrared Camera (NIRCam) on the JWST for GTO-1190 (PI: M. R. Meyer). This program includes three visits that occurred sequentially on 2023 March 1. Over all the visits, eight filters were used to obtain deep exposures of a single field in the core of NGC 2024, covering roughly 9.68 square arcminutes. This star-forming region is compact, young (<1 Myr), and nearby (∼400 pc); it contains massive stars; and it is highly extincted, with $A_V > 20$ magnitudes to many known sources (M. R. Meyer 1996; M. R. Meyer et al. 1997; M. Robberto et al. 2024), serving as a screen for many potential background contaminants while requiring deep exposures to detect free-floating objects down to sub-Jupiter masses. We observed this region with the F070W, F115W, F140M, F182M, F356W, F360M, F430M, and F444W filters, covering a broad range of wavelengths in order to detect brown dwarfs and free-floating planetary-mass objects. Four filters were used in the first and third visits, each comprised of three groups, five integrations, and seven total dithers with the standard subpixel dither type. The second visit used all eight filters and a sequence of three groups, six integrations, and three total dithers with the standard subpixel dither type. For each of the filters, we had 10 separate dithers and an effective exposure time of 6828.58 s or 7.6 hr of total science time.

## 3. Data Analysis

### 3.1. Data

In this Letter, we use the Stage 2 cal files and Stage 3 i2d files to perform our analyses. We obtained these data from the Mikulski Archive for Space Telescopes (MAST) at the Space Telescope Science Institute, and they can be accessed online.[17] The saturation limit of a single group using readout pattern SHALLOW2 is 16 mag in the F182M filter, corresponding to a 0.017 $M_\odot$ object with $A_V = 0$ or a 0.06 $M_\odot$ object with $A_V = 20$ at 400 pc, the estimated distance to NGC 2024 (J. E. Großschedl et al. 2018), using the 1 Myr isochrone of ATMO2020 (M. W. Phillips et al. 2020). The saturation limit of a single group in the F430M filter is 13.5 mag corresponding to 0.0485 $M_\odot$ with $A_V = 0$ for a 1 Myr source at 400 pc. Throughout the data, there are many saturated stars and high-mass brown dwarfs that are excluded from our analyses. We ignore any sources with at least one saturated pixel in the first group.

[17] doi:10.17909/vxaw-k902

### 3.2. Point-source Identification

The nebulosity within NGC 2024 extends throughout the entirety of our images in a variable, asymmetric structure with many filaments and knots; see Figure 1. When attempting to run common point-source detection algorithms suited for dense fields on these data (e.g., A. Dolphin 2016), many thousands of false-positive "point sources" are identified throughout the nebulosity of the cluster. Other false-positive detections are made in the extended wings of bright sources.

In order to combat this, we created a new process to filter out false positives and exclusively identify point sources. We briefly summarize the process below; see Appendix A for a thorough description of this routine. We background subtract Stage 2 data products. Then, we created subarrays of 32 × 32 pixels across the entire field of view and identified 5$\sigma$ detections within each subarray with DAOStarFinder (P. B. Stetson 1987), where $\sigma$ is the standard deviation within the subarray. We perform this subarray approach each of the 10 dither images for each filter. Sources are required to have the centroid within 0.5 pixels (15.5 mas in the SW channel and 31.5 mas in the LW channel) of the same detection in 8 of the 10 dithered images within the same filter. We also require each detection to be made within two or more filters, with an average centroid distance of 10 mas (SW channel) and 20 mas (LW channel) from the same source, comparable to the expected measurement uncertainties in the reference frame (M. De Furio et al. 2023). We remove any source within 10 pixels from saturated pixels. After this automated routine, we are left with 118 point-source candidates. Last, we evaluate the FWHM of each source. Sources are retained if they are measured with an FWHM < 1.25× the empirically stated FWHM[18] within two or more filters. This results in a final sample of 100 candidate sources. This automated approach ensures that our sample of point sources is not biased by any "by-eye" rejection and that our sensitivity is defined consistently across the entire field of view.

### 3.3. Photometry

In this section, we briefly summarize our approach to measuring photometry; see Appendix B for a more complete description. We used the PHOTMJSR value in the header of each Stage 3 file to convert from MJy/sr to counts/s. Then, we subtracted the mean sky background using an annulus with inner and outer radii of 0.″2 and 0.″3 in the shortwave channel and 0.″3 and 0.″4 in the longwave channel. Next, we used a circular aperture with radius equivalent to 0.″1 for the shortwave channel and 0.″2 for the longwave channel, to arrive at the total flux in the aperture, and divided by the corresponding encircled energy fraction for the filter in question, running from 70% to 78%. Finally, we calculates the Vega magnitude for each source using the zero points of each filter provided by STScI.[19]

[18] https://jwst-docs.stsci.edu/jwst-near-infrared-camera/nircam-performance/nircam-point-spread-functions

[19] https://jwst-docs.stsci.edu/files/182256933/224166043/1/1695068757137/NRC_ZPs_1126pmap.txt

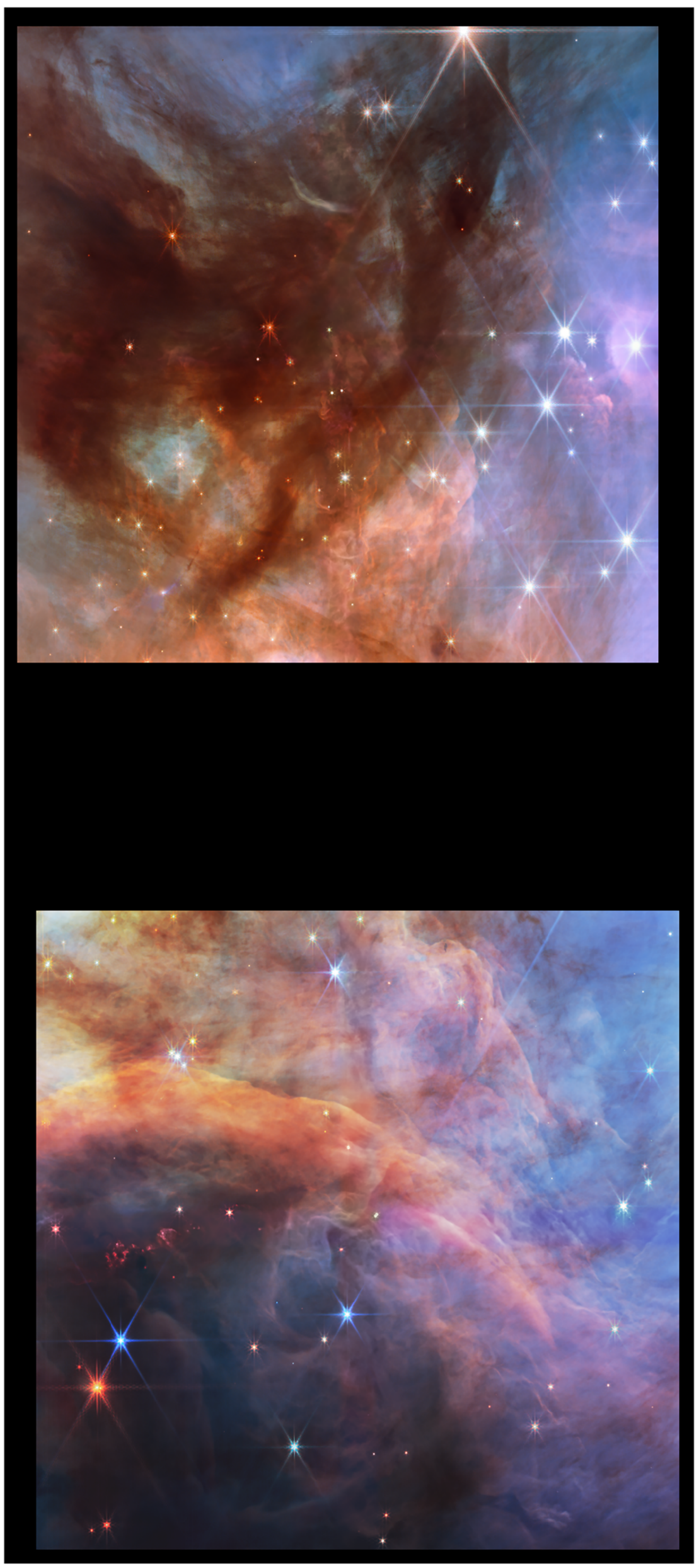

**Figure 1.** Color image of NGC 2024. Filters F115W and F140M are blue, F182M is green, F360M is orange, and F430M is red. Gaps in shortwave channel data are filled in using VISTA\VIRCAM data for reference. The F430M filter serves as the luminosity filter for the VISTA image, where the blue colors come from VISTA but the resolution comes from the F430M data. Shifting colors and balance adjustments are then made to fill in the gaps from the shortwave data. Credit: Alyssa Pagan, STScI.

In Figure 3, we present color–magnitude diagrams for sources that had photometry in the F182M and F430M filters. Importantly, there is a degeneracy between age, mass, and extinction that requires careful consideration to remove background field stars and brown dwarfs or extragalactic objects.

### *3.4. Background Object Contamination*

NGC 2024 has a prominent ridge in the core producing $A_V > 100$ mag, but the extinction is spatially variable, which allows background stars, brown dwarfs, and galaxies to appear as faint point sources. Due to the proximity of the cluster and the field of view of the data, we anticipate zero foreground objects that would resemble brown dwarf members. We evaluated the likelihood of each point source being a background contaminant by simulating a Galactic field population using the TRILEGAL tool (L. Girardi et al. 2005) and incorporating the Jaguar Mock Catalog (C. C. Williams et al. 2018), consistent with observations of the JADES program, to simulate the extragalactic population. For each point source, we estimate a maximum extinction if the source were truly background using the Herschel Gould Belt Survey (HGBS) molecular hydrogen column density maps (HGBS team 2020; see also P. André et al. 2010; P. Palmeirim et al. 2013) and the relation $N(\mathrm{H}_2) = 0.94 \times 10^{21} \times A_V$ [$\mathrm{cm}^{-2}\,\mathrm{mag}^{-1}$] (R. C. Bohlin et al. 1978; N. Schneider et al. 2022). We then project the simulated Galactic and extragalactic populations into color–magnitude space in many combinations of our filter sets; see Figure 2. Finally, for each point source, we calculate the expected number of background contaminants, given its color–magnitude and the estimated extinction at its location. For our sample of 100 point sources, we identified 48 likely cluster members and 52 likely background objects. Many faint, bluer point sources were consistent with background objects, similarly to the findings of K. L. Luhman (2024); see Figure 3. A more detailed discussion of this approach is included in Appendix C.

## 4. Mass Function Fitting

### *4.1. Sensitivity*

Importantly, our subarray point-source detection approach described in Section 3.2 allows us to obtain $5\sigma$ flux limits across the entire field of view in each filter. We calculate the fractional area of the field of view where sources are detectable above the limit using each subarray. We define our sensitivity to sources from 1% to 100% of the field of view, resulting in some defined degree of sensitivity from F430M = 13.5–23.6 mag and F182M − F430M = 0–11.2 mag; see Figure 4. Our flux limits correspond to a 0.5 $M_\mathrm{J}$ object with $A_V = 0$, a 1 $M_\mathrm{J}$ object with $A_V \sim 30$ mag, and a 2 $M_\mathrm{J}$ object with $A_V \sim 50$ mag (S. Wang & X. Chen 2019; M. W. Phillips et al. 2020). Brighter sources are ignored due to saturation that removes sources with F182M < 16 mag or F430M < 13.5 mag, which is most important for stars and high-mass brown dwarfs, which are not included in our analysis. In future work, we can improve the dynamic range of the data to higher-mass objects that are saturated by using the "Frame 0" images and by performing point-spread function fitting to extended wings of saturated sources. However, this population has been explored previously, and those higher-mass objects are outside the scope of this work.

### *4.2. Bayesian Analysis*

In order to fit a mass function to our data, we used a Bayesian framework to incorporate both the information from our likely cluster member detections and our sensitivity across the field of view, accounting for the incompleteness (see

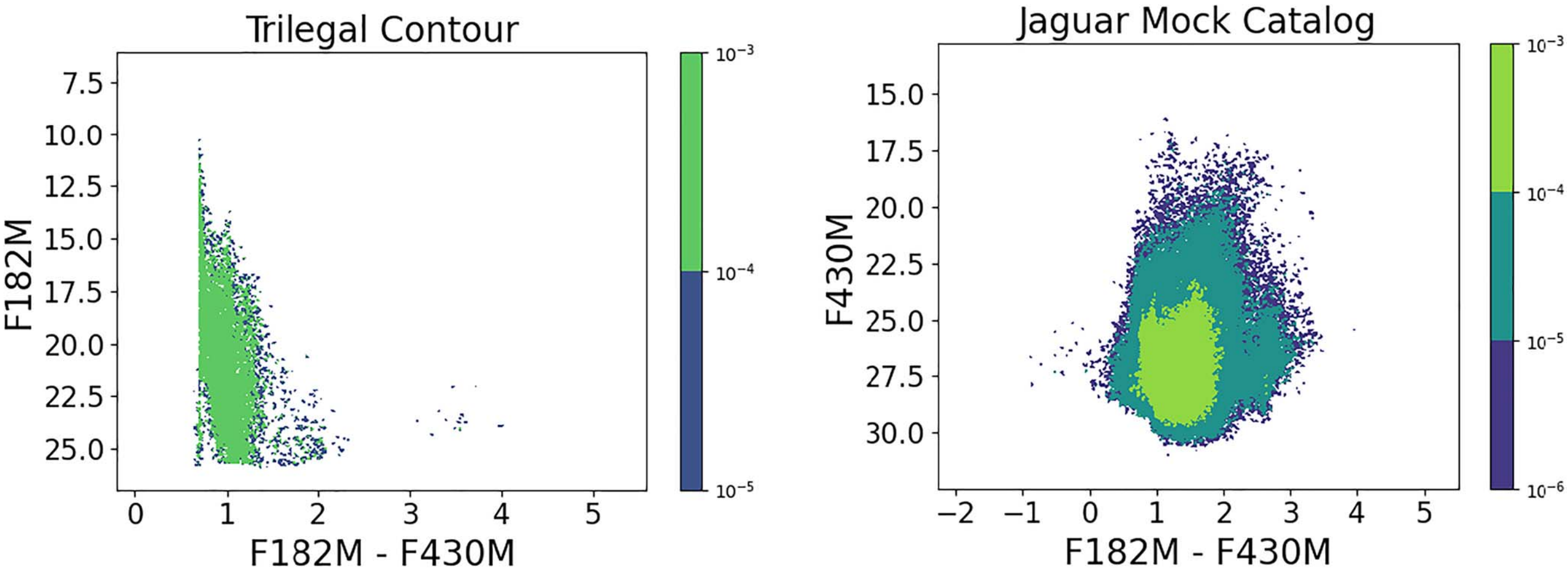

**Figure 2.** F430M vs. F182M − F430M diagram for the TRILEGAL (left) and JADES mock catalog (right) without any extinction applied. For all point-source detections, we applied the corresponding extinction from the HGBS to both of these distributions, calculated the number of expected contaminants given the observed color–magnitude of our targets, and normalized to the ratio of the areas of these catalogs to our field of view. All sources where the expected number of contaminants was <0.01 were classified as candidate cluster members.

Figure 4). First, we define a functional form for the mass function, such as a power law, similarly to previous work in the Galactic field (e.g., W. M. J. Best et al. 2024; J. D. Kirkpatrick et al. 2024):

$$N = C \int_{0.0005}^{0.062} m^{-\alpha} dm. \quad (1)$$

In this case, $N$ is the total number of objects across our full mass range and all extinctions $A_V = 0$–100 mag after correcting for incompleteness, and $\alpha$ is the index to the power law. Our bounds are 0.0005–0.062 $M_\odot$, the extent of the ATMO2020 atmospheric models for 1–2 Myr (M. W. Phillips et al. 2020), with $C$ as an integration coefficient. We analyze the F430M and F182M − F430M color–magnitude diagram to perform a homogeneous analysis for the detected population, which includes the largest number of objects in the sample and probes the lowest masses attainable with our survey 0.5 $M_J$. Of the 48 likely cluster members that were detected, 20 were not included in this analysis. Five were too faint to detect in the shortwave channel filters due to very high extinction, 10 were saturated in all the longwave channel filters and are likely stellar in mass, and the remaining five were unsaturated in the F430M filter but consistent with higher-mass brown dwarfs >60 $M_J$. We ignore these five additional sources because they are higher-mass and would require us to mix evolutionary models instead of consistently using the same model for the entire population. Importantly, this portion of the population has been thoroughly explored and modeled previously (e.g., M. Robberto et al. 2024) and is outside the scope of this work.

We utilize the Bayesian inference technique, PyMultiNest (J. Buchner et al. 2014), that performs the Nested Sampling Monte Carlo analysis (F. Feroz et al. 2009). For each fit to the data, PyMultiNest calculates the Bayesian evidence, a value we can use to compare to other model mass functions. We define the likelihood function with two components, the first of which is based on Poisson statistics and evaluates the model given our sensitivity. To calculate this, we generate a mass function from the sampled values of $\alpha$ and then sample that mass function to produce an artificial population of $10^3$ sources with random extinction = 0–100 mag. We convert the mass of each object into color–magnitude space using the ATMO2020 models, and apply the random extinction. Then, we evaluate the probability that each artificial source would be detected, given the sensitivity of our survey, and normalize the sampled number of expected detections ($N$) to the number of artificial generated sources ($n = 10^3$):

$$k = \sum_{i=1}^{n} p_i * \frac{N}{n}, \quad (2)$$

where $p_i$ is the probability a source would be detected given our sensitivity (colormap in Figure 4). Then we calculate the Poisson likelihood from the true number of detections in our sample ($d = 28$ substellar objects in F182M and F430M) and the evaluated value $k$:

$$\mathcal{L}_p = \frac{k^d e^{-k}}{d!}. \quad (3)$$

The second component of the likelihood calculation evaluates the probability of our true detections given the model. Following prescriptions in C. Fontanive et al. (2018) and M. De Furio et al. (2022a), we define a joint probability given our sensitivity (see Figure 4) multiplied by the mass function after it is mapped into color–magnitude space over $A_V = 0$–100 mag. Then, we evaluate the joint probability of each sampled mass function given the observed color–magnitude of our true detections, $p_j$. Finally, we combined these two calculations to arrive at the likelihood of each sampled mass function model:

$$\mathcal{L} = \mathcal{L}_p * \prod_{j=1}^{d} p_j. \quad (4)$$

For the single power-law case, we sample values of $N = [1, 100]$ and $\alpha = [-3, 3]$. From our sample of 28 detected substellar sources in the F182M and F430M filters, we derive 95% confidence intervals of $N = 43^{+17}_{-16}$ and $\alpha = 0.34^{+0.33}_{-0.37}$ over mass = 0.0005–0.062 $M_\odot$ and $A_V = 0$–100 mag (see Figure 5).

We also fit alternative mass function models (a single power law with a low-mass cutoff, a log-normal distribution, and a normal distribution) that return a difference in log-evidence <2 relative to the best fit of the single power law, or a probability

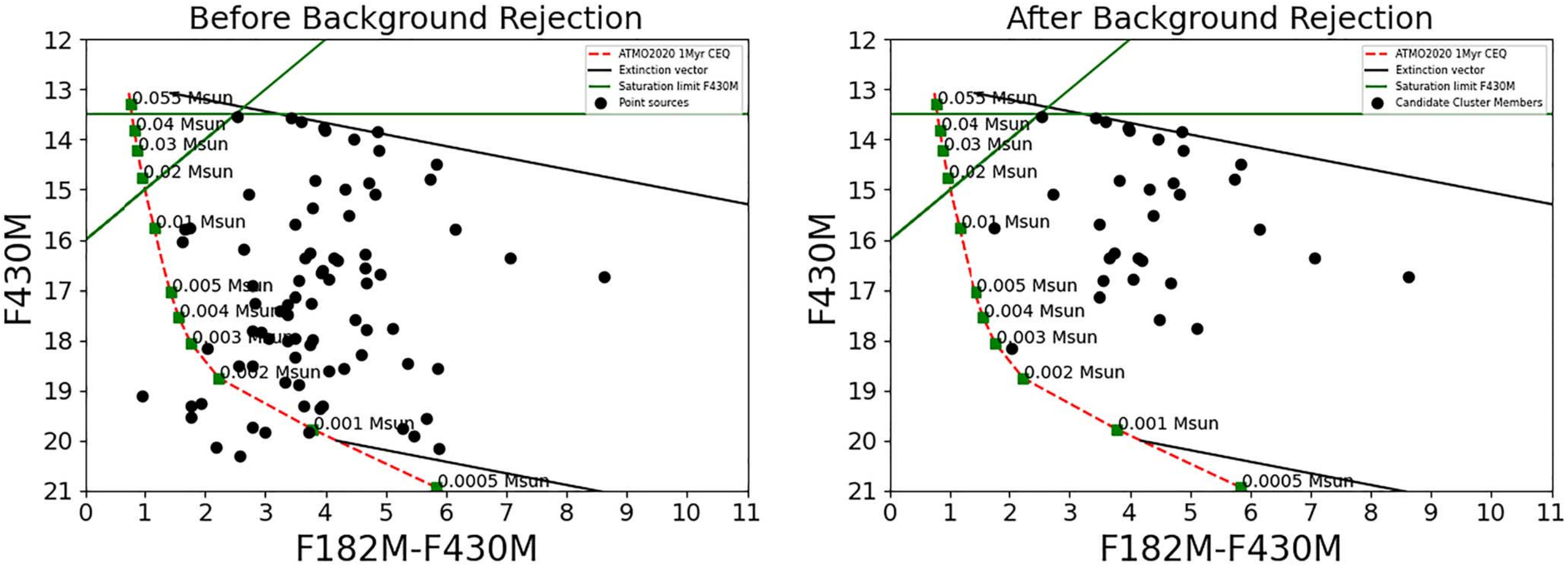

**Figure 3.** F430M vs. F182M − F430M before and after background rejection. Green lines are the saturation limits of the data with SHALLOW2 readout pattern. Black lines are extinction vectors (S. Wang & X. Chen 2019). Red line is ATMO2020 chemical equilibrium 1 Myr isochrone within various masses labeled in the green squares. Left: black circles are point sources identified within the data. Right: black circles are point sources within the data that have color–magnitude values that would indicate <0.01 background contaminants in our field of view, assuming the extinction from the HGBS.

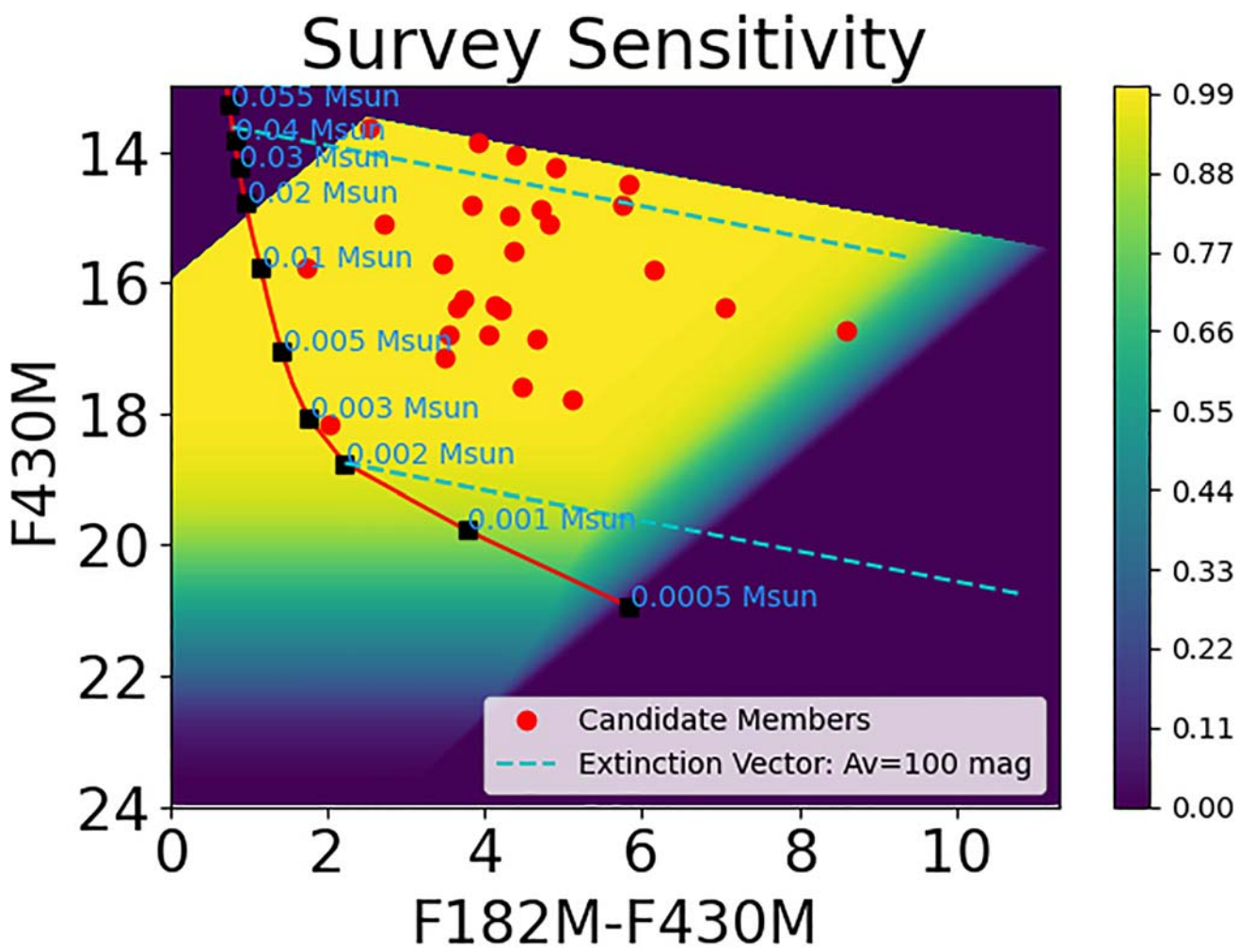

**Figure 4.** Sensitivity across the field of view of our NIRCam data in the F182M and F430M filters. Color bar shows the fractional area where we are sensitive to the specific color–magnitude combination. The purple background indicates no sensitivity, either due to saturation or being below our 5$\sigma$ limit.

<0.92 that these models are favored over the single power-law model (R. Trotta 2008).

Last, we fit a double power-law model and defined the mass function as

$$N = C_1 \int_{0.0005}^{m_{\rm BP}} m^{-\alpha_1} + C_2 \int_{m_{\rm BP}}^{0.062} m^{-\alpha_2}, \tag{5}$$

where $N$ is as above, $m_{\rm BP}$ is the breakpoint mass or where the two power laws differentiate, and $\alpha_1$ and $\alpha_2$ are the power-law indices for the low-mass portion and high-mass portion, respectively. For this model, we sample values of $N = [1, 100]$, $\alpha_1 = [-3, 3]$, $\alpha_2 = [-3, 3]$, and $m_{\rm BP}$ $[M_\odot] = [0.001, 0.061]$. We derive 95% confidence intervals of $N = 41^{+18}_{-17}$, $\alpha_1 = -1.03^{+1.07}_{-1.83}$, $\alpha_2 = 0.34^{+1.22}_{-1.63}$, and $m_{\rm BP} = 0.0120^{+0.0087}_{-0.0015}$ $[M_\odot]$ over mass = 0.0005–0.062 $M_\odot$ and $A_V$ = 0–100 mag; see Figure 5. Importantly, the difference in the log-evidence between this double power-law model and the single power-law model is 8.2, which indicates a probability >0.993 that the double power-law model is favored over the single power-law model (R. Trotta 2008). Among all the models we tested, the double power-law model with a breakpoint mass is highly favored to best resemble the results of our survey.

As a check on our analysis, we compared the single and double power-law models, generating the artificial population with a sampled extinction distribution derived from the 28 detections in Figure 4 instead of a uniform distribution from $A_V = 0$ to 100 mag. We found no change in the Bayesian evidence after applying different extinction distributions for either the single or double power-law models, and we found a slight increase in the power-law index for the low-mass portion of the mass function, consistent within the errors. We also compared the single and double power-law models with uniform $A_V$ distribution, placing hard minimum sensitivity cutoffs in our survey field of view at 80%, 90%, and 95% instead of the chosen full range of sensitivity (see Figure 4), and found a similar difference in the Bayesian evidence. The only change was the median of the low-mass power-law index ($\alpha_1$) that increased from −1.03 for the full sensitivity case to −0.89, −0.75, and −0.50 for the 80%, 90%, and 95% sensitivity cutoffs, respectively. This is due to the reduced ability to recover low-mass sources for higher-sensitivity cutoffs. Since our Bayesian approach accounts for the incompleteness of our survey, we chose to allow the full range of sensitivities to place constraints on the mass function. Finally, we reran all of our analyses, assuming different ages for the population. S. E. van Terwisga et al. (2020) identify an age spread of ∼0.5–1.5 Myr in NGC 2024. To model this potential age difference, we used the 1.6 and 2 Myr evolutionary models from ATMO2020, and find that all free parameters are well within 1$\sigma$ of the results from the 1 Myr evolutionary model. The derived Bayesian evidence between the models, the shape of the mass function, and the breakpoint mass are all consistent regardless of age assumed from 1 to 2 Myr.

## 5. Discussion

For our double power-law model, the indices for the low- and high-mass ends are within the respective 95% confidence intervals, due to the small number of detections in our sample. Additionally, the high-mass (∼12–62 $M_{\rm J}$) power-law index ($\alpha_2 = 0.34 + 1.22/-1.63$) is comparable to the single power-law index ($\alpha = 0.34 + 0.33/-0.37$) over 0.0005–0.062 $M_{\odot}$. The low-mass (∼0.5–12 $M_{\rm J}$) power-law index for our double power-law model has a 95% confidence interval of $\alpha_1 = [-2.86, 0.04]$, overlapping the 95% confidence interval of the power-law index for the single power-law model. We do not have a strong constraint on these power-law indices, due to the sample size; however, the Bayesian evidence strongly supports a model that rises from about 62 $M_{\rm J}$ to 12 $M_{\rm J}$ that then turns over and decreases down to 0.5 $M_{\rm J}$. This is the first evidence for a turnover in the mass function below 12 $M_{\rm J}$ in any stellar population, to our knowledge.

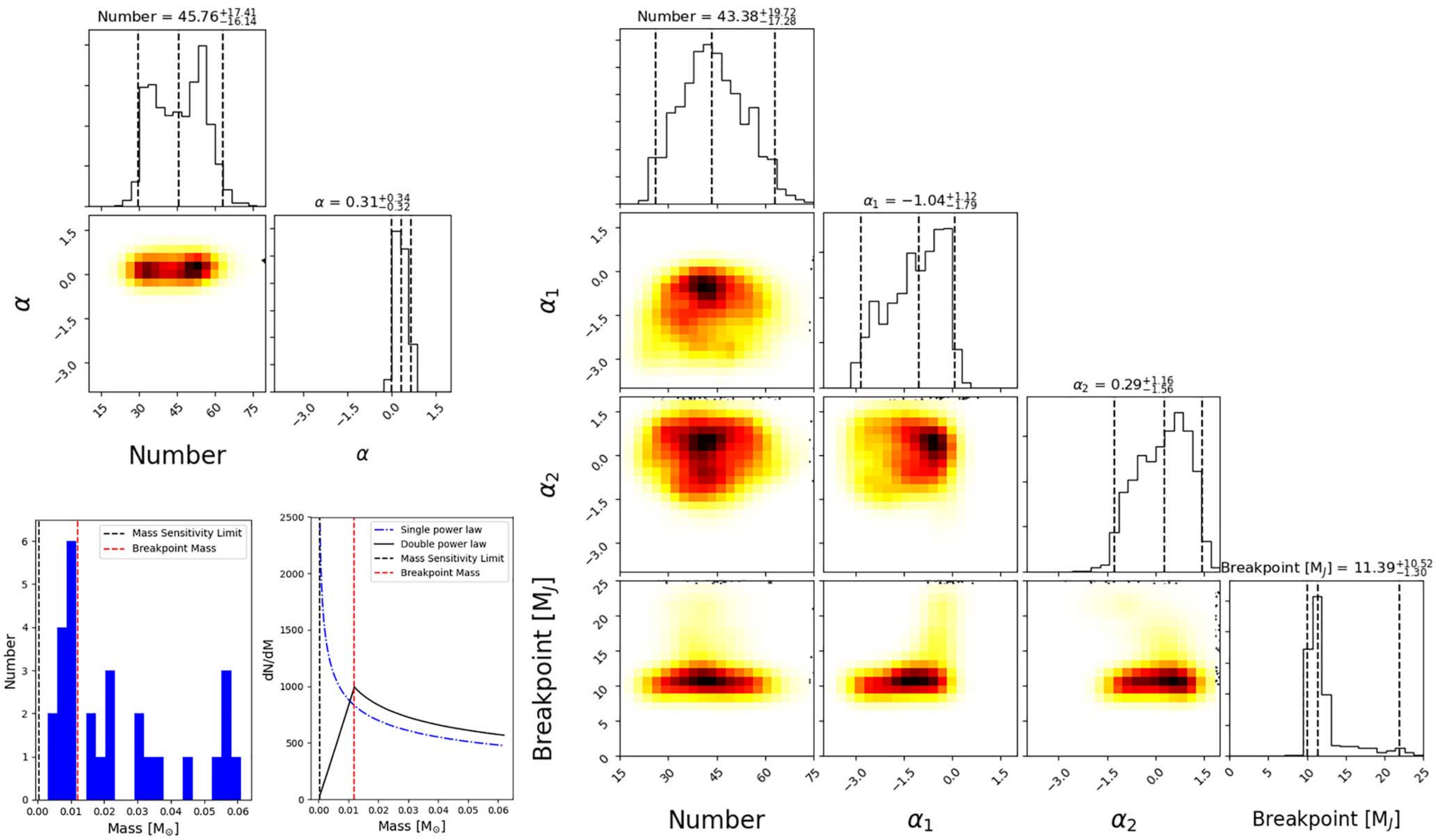

**Figure 5.** Top left: Corner plot for the single power-law model over 0.0005–0.062 $M_{\odot}$ and $A_V = 0$–100 mag. Number is the number of expected sources over the mass and extinction range after accounting for incompleteness of our sensitivity, and $\alpha$ is the power-law index. Right: Corner plot for the double power-law model where $\alpha_1$ is the power-law index from 0.0005 $M_{\odot}$ to the breakpoint mass and $\alpha_2$ is the power-law index from the breakpoint mass to 0.062 $M_{\odot}$. The breakpoint mass is the mass at which the two power laws diverge in units of $M_{\rm J}$. Bottom far left: Histogram with the estimated masses of the 28 likely cluster members detected and unsaturated in the F182M and F430M filters. Masses are derived from the 1 Myr ATMO2020 evolutionary model and their F182M − F430M color and F430M magnitude. The red dashed line is the breakpoint mass from the fit, and the black dashed line is sensitivity limit of our analysis. Bottom left: median values for the single (blue dotted–dashed line) and double (black solid line) power-law cases from our mass function fitting with the median value of the breakpoint mass (red dashed line) and the mass sensitivity limit of our analysis (black dashed line), 0.0005 $M_{\odot}$. The best-fit single power-law model predicts many Jupiter-mass objects while the best-fit double power-law model reflects a lack of objects down to 0.0005 $M_{\odot}$, as shown in our results.

With the results of our Bayesian analysis, we can compare to previous surveys in order to characterize the mass function below the theoretical limit of turbulent fragmentation, ∼2 $M_{\rm J}$ (D. F. A. Boyd & A. P. Whitworth 2005; A. P. Whitworth et al. 2024). Our single power-law model is consistent with the previous results in the Galactic field: $\alpha = 0.3 \pm 0.7$ over 0.01–0.08 $M_{\odot}$ (P. Kroupa 2001), $\alpha = 0.6 \pm 0.1$ for 0.01–0.05 $M_{\odot}$ (J. D. Kirkpatrick et al. 2021, 2024), and $\alpha = 0.8$ with a $3\sigma$ confidence interval of [0.2, 1.3] over masses 0.01–0.08 $M_{\odot}$ (P. Mróz et al. 2017). For our double power-law model, we directly compare the posterior probability distributions for the low-mass and high-mass regions to these other surveys by randomly sampling their value of $\alpha$ given their error bars. For each random sample, we then integrate the posterior distribution of $\alpha_1$ and $\alpha_2$ from that value to infinity, to arrive at the portion of the posterior that overlaps the single power-law models of other studies. Compared to early results in the Galactic field (P. Kroupa 2001), we find a mean integrated probability of 0.09 and 0.69 relative to the low-mass and high-mass power-law indices, respectively, i.e., no evidence for a difference. Examining more recent results in the field (J. D. Kirkpatrick et al. 2024), we find a mean integrated probability of 2.0e-4 and 0.61 for $\alpha_1$ and $\alpha_2$, respectively. Relative to microlensing studies (P. Mróz et al. 2017), we find a mean integrated probability of 1.6e-4 and 0.69 for $\alpha_1$ and $\alpha_2$, respectively. These results show strong evidence for a difference in the power law for the low-mass portion of the mass function and previous results in the field down to 0.01 $M_{\odot}$, supporting our key finding, i.e., a turnover in the mass function below ∼12 $M_{\rm J}$.

Other surveys in young, star-forming regions also probe the low-mass end of the mass function. Recent reviews (N. Bastian et al. 2010) report that most studies in star-forming regions identify a power-law index $\alpha \lesssim 0.5$ below the hydrogen-burning limit and down to ∼0.03 $M_{\odot}$, although studies at the time were not able to place strong constraints on $\alpha$. A study in Upper Sco (N. Miret-Roig et al. 2022) was sensitive down to ∼10 $M_{\rm J}$ and found results consistent with the Galactic field (J. D. Kirkpatrick et al. 2024). Due to their sensitivity limit near 10 $M_{\rm J}$, N. Miret-Roig et al. (2022) would not be able to identify a turnover in the mass function similar to our results. Other studies performed a meta-study of seven star-forming regions and used the ratio of stars to brown dwarfs (all >0.03 $M_{\odot}$) to compare to models of the mass function (M. Andersen et al. 2008). They found that their observations agree with IMF models and suggest a decrease in the mass function into the substellar regime (G. Chabrier 2003), although without the sensitivity to Jupiter-mass objects. A recent study in the Orion Nebula Cluster (ONC) with the Hubble Space Telescope found $\alpha = 0.6 \pm 0.06$ down to ∼5–7 $M_{\rm J}$ (M. Gennaro & M. Robberto 2020), providing 68%, 95%, and 99% errors. The high-mass end of our double power-law model is consistent with these results, with a mean integrated posterior probability of 0.62, but the low-mass portion of our double power-law model has a mean integrated probability of 3e–4, just as with the field population. Again, this comparison supports our identification of a break in the mass function down to Jupiter masses. M. Gennaro & M. Robberto (2020) fit a triple power-law model up to 1.4 $M_{\odot}$. It is likely that they did not identify a turnover in the mass function due to the lack of power in their detected population between 5 and 10 $M_{\rm J}$ relative to the rest of the stellar and substellar population. Further studies in the ONC with JWST (e.g., M. J. McCaughrean & S. G. Pearson 2023; S. G. Pearson & M. J. McCaughrean 2023) will allow us to define the mass function down to sub-Jupiter masses and search for environmental differences relative to our findings in NGC 2024.

Our observations cover the core of NGC 2024, potentially biasing results to the environmental dependencies of the star formation and evolution process, as previously shown in NGC 2024 for the stellar disk population (S. E. van Terwisga et al. 2020). It may be difficult to produce low-mass dense cores that will eventually become Jupiter-mass objects within the high stellar density regions of clusters or near ionizing sources, potentially requiring quiescent portions of the cluster for formation. Previous results in NGC 1333 and the ONC suggest that brown dwarf formation is environmentally dependent within a cluster, where brown dwarfs may form more often closer to other brown dwarfs as opposed to stellar cluster members (J. Greissl et al. 2007; M. Andersen et al. 2011). Other factors, such as episodic accretion, could impact the derived mass estimates for these objects but are typically rare, undefined at such low masses, and not expected to contribute large flux variations around 4 $\mu$m where our analysis relies. Even including this factor would likely not significantly impact the derived IMF shape statistical description, as it would contribute error to the mass estimates across the full population. Also, if extinction is a function of the masses of the objects, then our modeling of the mass function would be biased; however, there is no evidence for a mass-dependent extinction distribution in our data nor the literature. Additionally, unresolved binaries could cause the measured photometry and subsequent mass function modeling to overestimate the flux and mass of individual sources. Assuming a 15% companion frequency (e.g., C. Fontanive et al. 2018; M. De Furio et al. 2022b), four objects could be unresolved binaries. The low number of expected unresolved binaries leads us to conclude that the shape of the mass function is likely unchanged within the errors with a potential decrease in the breakpoint mass, conserving the turnover in the mass function. In GO-5409 (PI: M. De Furio), we will obtain NIRSpec/MSA prism spectroscopy of all likely substellar cluster members to determine membership and better estimate masses, and also obtain NIRCam photometry near the edge of the region to explore any environmental dependence.

Our observations in NGC 2024 are sensitive to 2 $M_{\rm J}$ objects with $A_V \sim 50$ mag and 1 $M_{\rm J}$ objects with $A_V \sim 30$ mag, albeit with no detections in our survey below ∼3 $M_{\rm J}$. These data suggest a potential limit of turbulent fragmentation near the theoretically derived value of ∼2 $M_{\rm J}$ (D. F. A. Boyd & A. P. Whitworth 2005; A. P. Whitworth et al. 2024). With these data, we can best model the mass function as a double power law with a breakpoint mass and identify the first turnover in the IMF within a stellar population down to 0.5 $M_{\rm J}$. These results have implications for planet ejection. The planet mass function rises to smaller masses (D. Suzuki et al. 2016), while our data indicate a falling mass function. The decrease in objects from 10 to 0.5 $M_{\rm J}$ indicates that ejected planets are not dominant among the observed population. Our work will serve as a benchmark for future programs devoted to exploring different Galactic star formation scenarios such as violent bursts in galaxies in the cores of giant ellipticals early in their formation that can produce a bottom-heavy IMF (P. G. van Dokkum & C. Conroy 2010). Similar populations may exist in the Galactic halo, potentially representative of early star formation in the Milky Way, that will be explored with surveys using JWST, the Rubin Observatory, and the Roman Space Telescope (C. Aganze et al. 2022; K. N. Hainline et al. 2024).

## 6. Conclusion

We performed a deep JWST/NIRCam imaging survey in the young (<1 Myr) embedded cluster NGC 2024 in order to characterize the mass function down to sub-Jupiter scales and search for free-floating planetary-mass objects. Our findings are as follows:

(1) We identified 48 likely substellar cluster members, many with colors consistent with estimated mass <10 $M_{\rm J}$, and the lowest with a mass estimate ∼3 $M_{\rm J}$. We are sensitive to lower masses, yet find no likely cluster members below 3 $M_{\rm J}$; this is suggestive of a potential limit to turbulent fragmentation near the theoretically derived value of ∼3 $M_{\rm J}$.

(2) We find that a double power-law model with a breakpoint mass best models our results compared to a single power-law model. We find the power-law index to the low-mass end of the mass function $\alpha_1 = -1.03^{+1.07}_{-1.83}$, the power-law index to the high-mass end of the mass function $\alpha_2 = 0.34^{+1.22}_{-1.63}$, and the breakpoint mass where the power laws diverge $m_{\rm BP} = 0.0120^{+0.0087}_{-0.0015}$ [$M_{\odot}$], over mass = 0.0005–0.062 $M_{\odot}$ and $A_V$ = 0–100 mag. This is the first indication of a turnover in the mass function in a stellar population to date.

## Acknowledgments

We would like to acknowledge the early work and contributions from Veenu Suri and Karl Gordon that led to the planning of this program. We would also like to acknowledge Adam Kraus for helpful discussions, especially on statistical processes and interpretations. We are grateful for support from NASA through the JWST NIRCam project through contract number NAS5-02105 (M. Rieke, University of Arizona, PI). This work is based on observations made with the NASA/ESA/CSA James Webb Space Telescope. The data were obtained from the Mikulski Archive for Space Telescopes at the Space Telescope Science Institute, which is operated by the Association of Universities for Research in Astronomy, Inc., under NASA contract NAS 5-03127 for JWST. These observations are associated with program No. 1190. This research has made use of data from the Herschel Gould Belt survey (HGBS) project (http://gouldbelt-herschel.cea.fr). The HGBS is a Herschel Key Programme jointly carried out by SPIRE Specialist Astronomy Group 3 (SAG 3), scientists of several institutes in the PACS Consortium (CEA Saclay, INAF-IFSI Rome and INAF-Arcetri, KU Leuven, and MPIA Heidelberg), and scientists of the Herschel Science Center (HSC). T.G. acknowledges support from the NASA Next Generation Space Telescope Flight Investigations program (now JWST) via WBS 411672.07.05.05.03.02. D.J. is supported by NRC Canada and by an NSERC Discovery Grant. C.M. is funded by the European Union (ERC, WANDA, 101039452). E.F. has been supported by project AYA2018-RTI-096188-B-I00 from the Spanish Agencia Estatal de Investigación and by Grant Agreement 101004719 of the EU project ORP. G.C. thanks the Swiss National Science Foundation for financial support under grant number P500PT_206785. M.D.F. is supported by an NSF Astronomy and Astrophysics Postdoctoral Fellowship under award AST-2303911. M.D.F. acknowledges S.F. Views and opinions expressed are however those of the author(s) only and do not necessarily reflect those of the European Union or the European Research Council Executive Agency. Neither the European Union nor the granting authority can be held responsible for them.

## Appendix A
## Point-source Detection

Many automated routines exist in order to detect point sources within a given image. They present ways to identify point sources and extract their astrometry and photometry within Hubble Space Telescope (HST) data (J. Anderson & I. R. King 2006; A. Dolphin 2016). There are also new applications of these methods to JWST data for the NIRCam and NIRISS instruments (M. Libralato et al. 2023; D. R. Weisz et al. 2024). These algorithms are designed to work well in crowded fields with many point sources, but were not specifically designed for data with a large amount of extended emission from nebulosity.

The nebulosity within NGC 2024 extends throughout the entirety of our images in a variable, asymmetric structure with many filaments and knots. When attempting to run common point-source detection algorithms on these data, many thousands of nonexistent "point sources" are identified throughout the nebulosity of the cluster. Other false-positive detections are made in the extended wings of bright sources.

In order to combat the issues identified with previous algorithms, we developed our own process to filter out semi-resolved structures within the cluster, image artifacts, and features of other bright point sources. We first took the Stage 2 data products from the MAST archive (cal) for each dither position and each filter, and estimated a background model using the MMMBackground function from Photutils. We model the background with a $30 \times 30$ pixel box size and a $5 \times 5$ pixel filter size. For each cal file from each dither position and each filter, we create this model and subtract it from our data to make background-subtracted images.

Simply defining some detectable flux above the background limit is not sufficient across the entire field of view, because of the significantly variable background. Therefore, we created small subarrays of $32 \times 32$ pixels across the entire field of view and then identified $5\sigma$ detections within each subarray, where $\sigma$ is the standard deviation within the subarray. We used the DAOStarFinder (P. B. Stetson 1987) function of Photutils in order to identify these candidate point sources, and included the empirically derived full width at half maximum (FWHM) as the input to the 2D Gaussian assumed for each filter. In order to avoid false negatives emanating from sources appearing on the edges of the subarrays, each successive subarray was created 16 pixels away in one direction from the previous subarray, in order to make sure that all point sources would be detected in at least one of our subarrays. For a given filter, we iterate this subarray detection approach for each dither position, one subarray at a time. If we make a $5\sigma$ detection within the subarray, to count this source as a point-source candidate, we then require the centroid of this detection defined by DAOStarFinder to be with 0.5 pixels (15.5 milliarcseconds [mas] in the SW channel and 31.5 mas in the LW channel) of the same detection in 8 of the 10 dithered images within the same filter, allowing for some error due to detector effects, cosmic rays, or low signal-to-noise producing larger uncertainties on the positions. We require that any point source is $>10$ pixels away from a saturated pixel; this is intended to remove false positives resulting from the wings of bright sources that saturate the detector. These processes filter out many detector effects like hot pixels from our list of possible point sources. This process also removes many false detections from the extended nebular emission, both because a $5\sigma$ detection above the local subarray background is much higher in these nebulous regions and because the centroid of a potential detection is not consistent throughout all the dither positions, due to its moderately spatially resolved extended emission. However, this approach does not filter out all of the detections of the extended wings of a bright point-spread function (PSF), as their centroids are conserved across all dither positions, due to the conserved extended structure of PSFs. We then require each detection to be made within two or more filters, with an average centroid distance 10 mas (SW channel) and 20 mas (LW channel) from the same source in another filter, in order to be classified as a candidate point source. Occasionally, some detections made on the extended wings of a bright PSF are still conserved because of the similar PSF structure between overlapping filters like F356W and F360M or F430M and F444W. After this automated routine, we are left with 118 point-source candidates that likely include some false positives from the extended wings of bright sources seen in similar positions in two filters.

To finalize our list of candidate point sources, we then evaluated the FWHM of each source within each filter. For each source, we performed a cubic spline in the $x$ and $y$ directions, beginning from the peak pixel of each dither position of the Stage 2 cal files. We then averaged the FWHM in both directions and over all dithers to get the final FWHM value within a filter. We accepted a source as a point source if we measured the FWHM $< 1.25\times$ the empirically stated FWHM within two or more filters, accommodating potential detector effects like the brighter/fatter effect for bright yet unsaturated sources as well as measurement error on very faint detections. This resulted in a final sample of 100 candidate sources. The 18 $5\sigma$ detections that were identified as having significantly larger FWHMs appear to be mostly extended wings of PSFs, knots of nebulosity within the cluster, and extended objects that are likely galaxies, given their presence in mostly red filters. This automated approach ensures that our sample of point sources is not dependent on manual or "by-eye" rejection and that our sensitivity is defined consistently across the entire field of view.

## Appendix B Photometry

To obtain photometry for each of the 100 sources in our sample, we first used the PHOTMJSR value in the header of each Stage 3 i2d file to convert the units of the data from MJy/sr to counts/s. Then, we subtracted the mean sky background using an annulus with inner and outer radii of $0\farcs2$ and $0\farcs3$ in the shortwave channel and $0\farcs3$ and $0\farcs4$ in the longwave channel. Next, we used a circular aperture with radius equivalent to $0\farcs1$ for the shortwave channel and $0\farcs2$ for the longwave channel to arrive at the total flux in the aperture, and divided by the corresponding encircled energy fraction for the filter in question, running from 70% to 78%. Finally, we calculate the Vega magnitude for each source using the zero points of each filter provided by STScI.

For every source that was detected in two or more filters, we ran an additional check throughout all filters to determine whether the source was detected. Occasionally, easily identifiable sources in the red filters were not detected in F115W or F140M, even though they were very bright yet unsaturated. The severe undersampling in F115W and detector effects like charge migration and the brighter–fatter effect will significantly distort the core of the PSF even if it is not saturated. This caused our initial algorithm to skip over bright sources in the blue filters, due to them exhibiting a significantly different FWHM than the empirically derived relation. Therefore, this extra step is necessary to make sure all possible photometry is obtained for all detected point sources.

In Figure 3, we present color–magnitude diagrams for sources that had photometry in the F182M and F430M filters. We plot the calculated photometry compared to 1 Myr isochrones from the ATMO2020 chemical equilibrium models with extinction vectors derived for JWST filters (S. Wang & X. Chen 2019). The saturation limits of each filter are shown, corresponding to substellar objects even through many orders of magnitude in extinction. Importantly, there is a clear degeneracy between age, mass, and extinction that makes it difficult to identify candidate members purely based on their colors. Background field stars and brown dwarfs or extragalactic objects with high extinction can be confused with lower-mass brown dwarf members with less extinction.

## Appendix C Background Rejection

NGC 2024 has a prominent ridge in the core producing $A_V > 100$ mag. However, the extinction is spatially variable, allowing for the potential for background stars and galaxies to shine through and appear as faint point sources. We used the Herschel Gould Belt Survey (HGBS) $H_2$ column density maps (P. André et al. 2010; P. Palmeirim et al. 2013) to estimate the visual extinction expected at each point within our field of view, although the pixel size of Herschel is $3\farcs0$ compared to the $0\farcs063$ pixel of the longwave channel. Across the $\sim 9\farcs68$ field of view, we record $A_V$ values from 8 to 220 mag using the relation $N(\mathrm{H2}) = 0.94 \times 10^{21} \times A_V$ [cm$^{-2}$ mag$^{-1}$] (R. C. Bohlin et al. 1978; N. Schneider et al. 2022). In this region, there is significant ionization that can result in excess free–free emission. This free–free emission adds to the flux recorded by Herschel, leading to overestimates of extinction at the highest values. Therefore, extinctions as high as 220 mag may be nonphysical, but the spatial dependence of extinction is clear.

Two possible background contaminants exist for our sample: Galactic field stars/brown dwarfs and extragalactic objects. We simulated a Galactic field population using the TRILEGAL tool (L. Girardi et al. 2005). We input the coordinates for NGC 2024 and output a photometry table for the JWST/NIRCam filters in question using the default TRILEGAL prescriptions. Additionally, we downloaded the Jaguar Mock Catalog that simulates the expected distribution of galaxies for many JWST instruments and filters (C. C. Williams et al. 2018). For each catalog, we produced a grid in color–magnitude space in equally spaced bins of $<0.1$ mag, then determined the distribution of galaxies and field objects expected within the full range of color–magnitude space, and normalized the distribution (see Figure 2). If a source were background, it would be consistent with these distributions of Galactic and extragalactic sources after the estimated extinction was applied. For each source in question, we have an estimated extinction from the HGBS data as well as measured photometry with errors. We first produce a normalized Gaussian distribution of the source color–magnitude given the observed photometry and measured errors calculated from either the error arrays or assumed to be 3%, whichever is higher. An assumption of 3% errors is based on the continuous calibration errors that appear to change for many filters and detectors throughout the run of JWST. We then apply the expected extinction at that same position to the background source distributions. We then calculate the amount of overlap between the two distributions, summing the following metric over all color–magnitude space: $\sqrt{P1 * P2}$. This value tells us the fraction of overlap our source has with the background object distribution. Then, we multiply by the number of sources in each catalog and the ratios of the field of view in each catalog relative to our observations, to arrive at the total expected number of background objects within our field of view with the estimated extinction and the color–magnitude of our target. We performed this analysis for these combinations of color–magnitude: F115W − F182M versus F182M, F140M − F182M versus F182M, F115W − F430M versus F430M, F140M − F430M versus F430M, F182M − F430M versus F430M, and F360M − F430M versus F430M. For our sample of 100 objects, if the expected number of contaminants was $<0.01$ in one of the combinations of color–magnitude, we classified it as a likely cluster member. We identified 48 objects that survived the rejection process as likely cluster members, with 52 classified as background objects and the expectation of $<1$ false-positive result. All of our sources except for one were either very likely to be a background contaminant or very unlikely to be a background contaminant in multiple combinations of color–magnitude, highlighting the robustness of this approach.

## ORCID iDs

Matthew De Furio https://orcid.org/0000-0003-1863-4960
Michael R. Meyer https://orcid.org/0000-0003-1227-3084
Thomas Greene https://orcid.org/0000-0002-8963-8056
Klaus Hodapp https://orcid.org/0000-0003-0786-2140
Doug Johnstone https://orcid.org/0000-0002-6773-459X
Jarron Leisenring https://orcid.org/0000-0002-0834-6140
Marcia Rieke https://orcid.org/0000-0002-7893-6170
Massimo Robberto https://orcid.org/0000-0002-9573-3199
Thomas Roellig https://orcid.org/0000-0002-6730-5410
Gabriele Cugno https://orcid.org/0000-0001-7255-3251
Eleonora Fiorellino https://orcid.org/0000-0002-5261-6216
Carlo F. Manara https://orcid.org/0000-0003-3562-262X
Sierk van Terwisga https://orcid.org/0000-0002-1284-5831

## References

Aganze, C., Burgasser, A. J., Malkan, M., et al. 2022, ApJ, 924, 114
Anderson, J., & King, I. R. 2006, Instrument Science Report ACS 2006-01,
Andersen, M., Meyer, M. R., Greissl, J., & Aversa, A. 2008, ApJL, 683, L183
Andersen, M., Meyer, M. R., Robberto, M., Bergeron, L. E., & Reid, N. 2011, A&A, 534, A10
André, P., Men'shchikov, A., Bontemps, S., et al. 2010, A&A, 518, L102
Bastian, N., Covey, K. R., & Meyer, M. R. 2010, ARA&A, 48, 339
Best, W. M. J., Sanghi, A., Liu, M. C., Magnier, E. A., & Dupuy, T. J. 2024, ApJ, 967, 115
Bohlin, R. C., Savage, B. D., & Drake, J. F. 1978, ApJ, 224, 132
Boyd, D. F. A., & Whitworth, A. P. 2005, A&A, 430, 1059
Buchner, J., Georgakakis, A., Nandra, K., et al. 2014, A&A, 564, A125
Chabrier, G. 2003, PASP, 115, 763
De Furio, M., Lew, B., Beichman, C., et al. 2023, ApJ, 948, 92
De Furio, M., Liu, C., Meyer, M. R., et al. 2022a, ApJ, 941, 161
De Furio, M., Meyer, M. R., Reiter, M., et al. 2022b, ApJ, 925, 112
Dolphin, A. 2016, DOLPHOT: Stellar photometry, Astrophysics Source Code Library, ascl:1608.013
Feroz, F., Hobson, M. P., & Bridges, M. 2009, MNRAS, 398, 1601
Fontanive, C., Biller, B., Bonavita, M., & Allers, K. 2018, MNRAS, 479, 2702
Gennaro, M., & Robberto, M. 2020, ApJ, 896, 80
Girardi, L., Groenewegen, M. A. T., Hatziminaoglou, E., & da Costa, L. 2005, A&A, 436, 895
Greissl, J., Meyer, M. R., Wilking, B. A., et al. 2007, AJ, 133, 1321
Großschedl, J. E., Alves, J., Meingast, S., & Hasenberger, B. 2018, arXiv:1812.08024
Hainline, K. N., Helton, J. M., Johnson, B. D., et al. 2024, ApJ, 964, 66
HGBS team 2020, Herschel Gould Belt Survey, NASA IPAC DataSet, IRSA72, doi:10.26131/IRSA72
Kirkpatrick, J. D., Gelino, C. R., Faherty, J. K., et al. 2021, ApJS, 253, 7
Kirkpatrick, J. D., Marocco, F., Gelino, C. R., et al. 2024, ApJS, 271, 55
Kroupa, P. 2001, MNRAS, 322, 231
Langeveld, A. B., Scholz, A., Mužić, K., et al. 2024, AJ, 168, 179
Libralato, M., Bellini, A., van der Marel, R. P., et al. 2023, ApJ, 950, 101
Luhman, K. L. 2024, AJ, 168, 230
Luhman, K. L., Alves de Oliveira, C., Baraffe, I., et al. 2024, AJ, 167, 19
McCaughrean, M. J., & Pearson, S. G. 2023, arXiv:2310.03552
Meyer, M. R. 1996, PhD thesis, Max-Planck-Institute for Astronomy, Heidelberg
Meyer, M. R., Calvet, N., & Hillenbrand, L. A. 1997, AJ, 114, 288
Miret-Roig, N., Bouy, H., Raymond, S. N., et al. 2022, NatAs, 6, 89
Moraux, E., Kroupa, P., & Bouvier, J. 2004, A&A, 426, 75
Mróz, P., Udalski, A., Skowron, J., et al. 2017, Natur, 548, 183
Palmeirim, P., André, P., Kirk, J., et al. 2013, A&A, 550, A38
Pearson, S. G., & McCaughrean, M. J. 2023, arXiv:2310.01231
Phillips, M. W., Tremblin, P., Baraffe, I., et al. 2020, A&A, 637, A38
Rieke, M. J., Kelly, D. M., Misselt, K., et al. 2023, PASP, 135, 028001
Robberto, M., Gennaro, M., Da Rio, N., et al. 2024, ApJ, 960, 49
Salpeter, E. E. 1955, ApJ, 121, 161
Schneider, N., Ossenkopf-Okada, V., Clarke, S., et al. 2022, A&A, 666, A165
Stetson, P. B. 1987, PASP, 99, 191
Suzuki, D., Bennett, D. P., Sumi, T., et al. 2016, ApJ, 833, 145
Trotta, R. 2008, ConPh, 49, 71
van Dokkum, P. G., & Conroy, C. 2010, Natur, 468, 940
van Terwisga, S. E., van Dishoeck, E. F., Mann, R. K., et al. 2020, A&A, 640, A27
Wang, S., & Chen, X. 2019, ApJ, 877, 116
Weisz, D. R., Dolphin, A. E., Savino, A., et al. 2024, ApJS, 271, 47
Whitworth, A. P., Priestley, F. D., Wünsch, R., & Palouš, J. 2024, MNRAS, 529, 3712
Williams, C. C., Curtis-Lake, E., Hainline, K. N., et al. 2018, ApJS, 236, 33